\documentclass[proof]{WileyASNA-v1}

\articletype{Article Type}%

\received{}
\revised{}
\accepted{}

\begin{document}

\title{Hyperonic Neutron Star Matter in Light of GW170817}

\author[1]{William M. Spinella*}

\author[2,3]{Fridolin Weber}

\authormark{William M. Spinella \textsc{et al}}

\address[1]{\orgdiv{Department of Sciences}, \orgname{Wentworth Institute of Technology}, \orgaddress{\state{Massachussetts}, \country{United States}}}

\address[2]{\orgdiv{Department of Physics}, \orgname{San Diego State University}, \orgaddress{\state{California}, \country{United States}}}

\address[3]{\orgdiv{Center for Astrophysics and Space Sciences}, \orgname{University of California San Diego}, \orgaddress{\state{California}, \country{United States}}}

\corres{*\email{spinellaw@wit.edu}}


\abstract{Since 2013 the mass of pulsar PSR J0348+0432 ($M = 2.01\,M_{\odot}$)
has provided a tight constraint on neutron star equation of state.  However, a
number of different analyses of the recently detected BNS 
merger (GW170817) point to a maximum neutron star mass around $2.16\,M_{\odot}$.
In addition, a recent study determined the mass of the millisecond pulsar
PSR J2215+5135 to be $2.27^{+0.17}_{-0.15}\,M_{\odot}$. In this work we
investigate the presence of hyperons in neutron star matter in light of
these new mass measurements using equations of state calculated
in the relativistic mean-field approximation.
Particular attention is paid to the use of the available empirical data
from the study of hypernuclei and that of the SU(3) symmetry relations
in fixing the meson-hyperon coupling constants.  We find that
hyperonic equations of state with reasonable choices for the meson-hyperon
coupling constants can satisfy these new mass constraints, with 
hyperons potentially accounting for more than 10\% of the baryons in
the core of a neutron star.}




\maketitle

\footnotetext{\textbf{Abbreviations:} 
BNS, binary neutron star;
EoS, equation of state;
NS, neutron star;
RMF, relativistic mean-field;
SNM, symmetric nuclear matter}


\section{Introduction}\label{introduction}

Published mass measurements of PSR J1614-2230 in 2010 and PSR J0348+0432
in 2013 indicate that neutron stars (NSs) can have masses as large as twice
the mass of our sun.
Including hyperons in the modeled composition of NS matter increases
the baryonic degrees of freedom softening the NS equation of state
(EoS), the degree of which depends on the choice of meson-hyperon coupling
constants in Relativistic Mean-Field (RMF) models.
This softening often results in a calculated maximum NS mass that is
inconsistent with the aforementioned observations, giving rise to the
so-called ``hyperon puzzle."
Common resolutions of the hyperon puzzle involve stiffening the EoS by
introducing repulsive hyperon-nucleon and hyperon-hyperon interactions and
adjusting the meson-hyperon coupling constants in various EoS models.
However, analyses of the recent multi-messenger observation of a binary neutron
star (BNS) merger suggest a maximum NS mass significantly higher than
two solar masses, as does the recently published measurement of the mass of PSR
J2215+5135, reinvigorating the hyperon puzzle. 
Therefore, we seek to determine whether or not a reasonably constrained
hyperonic NS matter EoS can produce a maximum mass consistent
with these new constraints.

In this work we compute the EoS of NS matter including the full baryon octet, using
a RMF model consistent with constraints on the properties of the isospin-symmetric
nuclear matter (SNM) EoS including the nuclear incompressibility and the isospin
asymmetry energy and slope.
EoS stiffening due to the inclusion of hyperon-hyperon interactions, mediated via 
strange-scalar ($\sigma^*$) and strange-vector ($\phi$)
mesons, is examined by calculating NS properties for different
combinations of mesons.
Scalar meson-hyperon coupling constants are fit to empirical data on
hypernuclei where available, while the vector meson-hyperon couplings
are fixed by the SU(3) symmetry.
By varying SU(3) coupling parameters we traverse the entire vector
meson-hyperon coupling space and calculate NS properties including the maximum
mass, strangeness fraction, and hyperon fraction.
We determine the couplings that are consistent with the mass constraints
provided by PSR J0348+0432 and analyses of the detected gravitational waves
(GW170817) and gamma-ray burst (GRB 170817A) produced by the observed BNS
merger.

This work is organized as follows.
In Sec. \ref{sec:equation-of-state} we
introduce the relativistic mean-field model and parameterization of the
EoS, as well as the methods used for determining the meson-hyperon coupling
constants.
In Sec. \ref{sec:gw170817} we discuss additional constraints on the EoS
provided by analyses of GW170817 and GRB 170817A.
Our results, including calculations of NS properties for the vector
meson-hyperon coupling space, are presented in Sec. \ref{sec:results}.
Finally, we provide a summary of this work in Sec. \ref{sec:summary}.


\section{Equation of State}\label{sec:equation-of-state}

\begin{figure}[t]
\centerline{\includegraphics[width=0.5\textwidth]{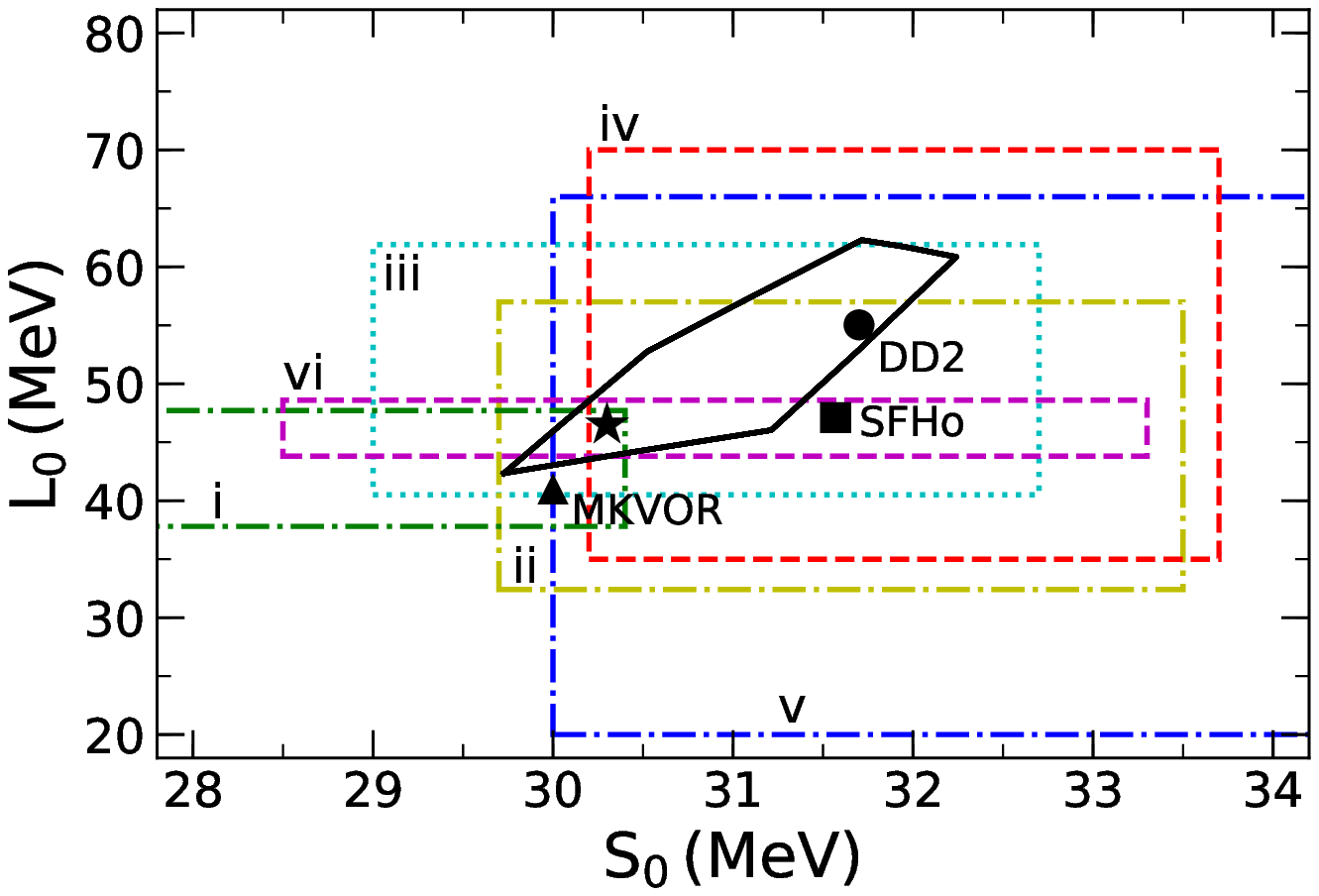}}
\caption{Symmetry energy and slope of the SNM parameterization at
saturation density with the following constraints from the literature:
(i) \cite{Hagen2015},
(ii) \cite{Hebeler2013},
(iii) \cite{Lattimer2014}, 
(iv) \cite{Danielewicz2014},
(v) \cite{Roca-Maza2015},
(vi) \cite{Birkhan2017}.
The irregular black outline represents the accepted region determined from
various experimental and theoretical constraints compiled in Fig. 9 of
\cite{Tews2017}. The star marker indicates the
values for the parameterization used in this work.}
\label{fig:symmetry-energy}
\end{figure}

In this work we calculate EoSs of cold, nonrotating NSs using the nonlinear
relativistic mean-field (RMF)
approximation, modeling baryon-baryon interactions in terms of scalar
($\sigma,\sigma^*$), vector ($\omega,\phi$), and isovector
($\rho$) meson fields, and represented by the following Lagrangian,
\begin{eqnarray} \label{eq:Lagrangian}
  \mathcal{L} &=& \sum\limits_B \overline\psi_B\bigl[\gamma_{\mu}
	(i\partial^{\mu}-g_{\omega B}\omega^{\mu}-g_{\phi B}\phi^{\mu}-\tfrac{1}{2}g_{\rho B}\boldsymbol{\tau}\cdot
	\boldsymbol{\rho}^{\mu})\nonumber\\
  &&-(m_B-g_{\sigma B}\sigma-g_{\sigma^* B}\sigma^*)\bigr]\psi_B\nonumber\\
	&&+\tfrac{1}{2}\left(\partial_{\mu}\sigma\partial^{\mu}\sigma
	-m^2_{\sigma}\sigma^2\right)\nonumber\\
	&&-\tfrac{1}{3}b_{\sigma}m_n\left(g_{\sigma N}\sigma\right)^3
	-\tfrac{1}{4}c_{\sigma}\left(g_{\sigma N}\sigma\right)^4\nonumber\\
	&&-\tfrac{1}{4}\omega_{\mu\nu}\omega^{\mu\nu}
	+\tfrac{1}{2}m^2_{\omega}\omega_{\mu}\omega^{\mu}\nonumber\\
	&& -\tfrac{1}{4}\boldsymbol{\rho}_{\mu\nu}\cdot\boldsymbol{\rho}^{\mu\nu}\nonumber
	+\tfrac{1}{2}m^2_{\rho}\boldsymbol{\rho}_{\mu}\cdot\boldsymbol{\rho}^{\mu}\\
	&&-\tfrac{1}{4}\phi^{\mu\nu}\phi_{\mu\nu}+\tfrac{1}{2}m^2_{\phi}\phi_{\mu}
	\phi^{\mu}\nonumber\\
	&&+\tfrac{1}{2}\left(\partial_{\mu} \sigma^*\partial^{\mu}\sigma^*
	-m^2_{\sigma^*}\sigma^{*2}\right)\nonumber\\
&&+\sum_{\lambda}\overline\psi_{\lambda}\bigl(i\gamma_{\mu}\partial^{\mu}
				-m_\lambda\bigr)\psi_{\lambda}\,.
\end{eqnarray}
The EoS is parameterized to reproduce the following properties of SNM at a
saturation density of $n_0 = 0.150$ fm$^{-3}$: the energy per nucleon
$E_0/N = -16.0$ MeV, nuclear incompressibility $K_0 = 250.0$ MeV, effective nucleon
mass $m^*/m_N = 0.70$, isospin asymmetry energy $S_0 = 30.3$ MeV, and slope of
the asymmetry energy $L_0 = 46.5$ MeV.
The parameterization is specifically tailored for consistency with a number of
simultaneous constraints on the asymmetry energy ($S_0$) and slope of the asymmetry
energy ($L_0$) at $n_0$ as shown in Figure \ref{fig:symmetry-energy}.
In order to satisfy the constraints on $L_0$ the isovector-vector meson-baryon coupling
constants are taken to be density-dependent with a functional density-dependence
given by \cite{Typel1999,Drago2014a}
\begin{equation}
  g_{\rho B}(n) = g_{\rho B}(n_0)\,\mathrm{exp}\!\left[-a_{\rho}\left(n/n_0-1\right)
  \right]\,,
\end{equation}
with the parameter $a_{\rho}$ fit to $L_0$ at $n_0$.

\begin{figure}[t]
\centerline{\includegraphics[width=0.50\textwidth]{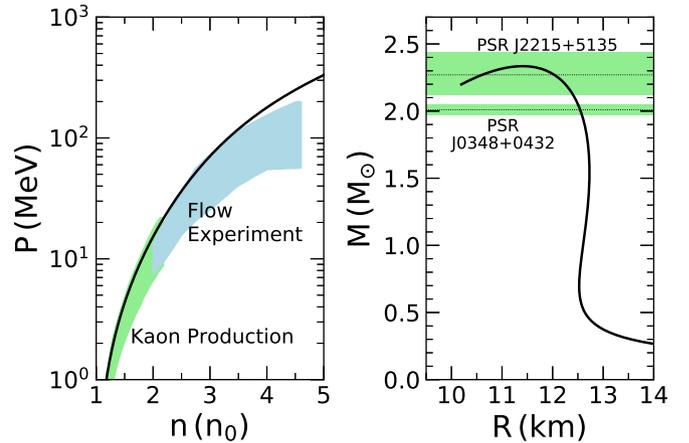}}
  \caption{ EoS of SNM (left panel) and mass-radius relation for a purely
  nucleonic EoS (right panel), both including constraints discussed in the text.}
\label{fig:eos-mass-radius}
\end{figure}

The EoS of SNM and mass-radius relation for the purely nucleonic EoS ($npe\mu$) are
given in Figure \ref{fig:eos-mass-radius}.
The low-density EoS is sufficiently soft to satisfy the constraint from kaon
production and produce a relatively small canonical radius,
while the high-density EoS is too stiff to fully satisfy the elliptic flow
constraint but produces a high maximum mass ($M_{\mathrm{max}}=2.33\,M_{\odot}$)
consistent with the masses of PSR J0348+0432 ($M=2.01\pm0.04\,M_{\odot}$)
and PSR J2215+5135 ($M=2.27^{+0.17}_{-0.15}\,M_{\odot}$)
\cite{Fuchs2006,Lynch2009,Antoniadis2013,Linares2018,Danielewicz2002}.
Note that outer and inner crust EoSs are appended to the low density EoS
\cite{Douchin2001,Haensel1994}.


\subsection{Meson-Hyperon Coupling Constants} \label{sec:coupling-constants}

Inclusion of the hyperonic degrees of freedom softens the neutron
star EoS, and the corresponding reduction of the maximum mass is extremely
sensitive to the choice of meson-hyperon coupling constants.
It is therefore essential that the coupling constants be fixed using
empirical data wherever possible.

The scalar meson-hyperon coupling constants $g_{\sigma Y}$ and $g_{\sigma^* Y}$
can be fit to hyperon single-particle potentials and self-potentials derived from
the available empirical data on hypernuclei, but first the vector meson-hyperon
couplings $g_{\omega Y}$ and $g_{\phi Y}$ must be specified.
In SU(3) symmetry the vector couplings can be written in terms of three parameters:
the mixing angle $\theta_V$, the $F/(F+D)$ ratio $\alpha_V$, and the
octet-singlet coupling ratio $z$ \cite{Dover1984}. 
The most commonly used values for these parameters are $\theta_V=35.26^{\circ}$, 
$\alpha_V=1$, and $z=1/\sqrt{6}$, corresponding to SU(6) symmetry.
However, sophisticated baryon-baryon interaction models such as the
Nijmegen extended-soft-core model (ESC08) predict much lower values of the
coupling ratio $z$ ($z_{\mathrm{ESC08}} \approx 0.195$) that lead to stiffer
EoSs and much higher maximum masses than those constructed using SU(6) symmetry
\cite{Rijken2010}.
Rather than calculate a single EoS constructed from fixed values of $\alpha_V$
and $z$, we will instead calculate NS maximum masses, strangeness
fractions, and hyperon fractions for the $z$ parameter space with $\alpha_V=1$,
and for the entire $\alpha_V$-$z$ parameter space.
Note that the $\phi$ meson couples to the nucleons if $z \ne 1/\sqrt{6}$,
and as a result $g_{\omega N}$ must be recalculated to restore the saturation
properties of the RMF parameterization.

With the vector meson-hyperon couplings specified, the scalar couplings are
set to reproduce empirical hyperon single-particle potentials at saturation,
$U_{Y}^{~\!(N)}(n_0)$, using the following,
\begin{equation} \label{eq:hyperon-nucleon-potential}
  U_{Y}^{~\!(N)}(n_0) = 
	  g_{\omega Y} \bar\omega_0
  + g_{\phi Y} \bar\phi_0
  - g_{\sigma Y} \bar\sigma_0\,.
\end{equation}
In this work we employ the following hyperon potentials:
$U_{\Lambda}^{(N)}(n_0) = -28$ MeV, $U_{\Sigma}^{(N)}(n_0) = +30$ MeV,
and $U_{\Xi}^{(N)}(n_0) = -14$ MeV.
If the strange-scalar field $\sigma^*$ is included, the strange-scalar
meson-$\Lambda$ coupling constant $g_{\sigma^*\Lambda}$ is set to reproduce a
saturation self-potential of $U_{\Lambda}^{(\Lambda)}(n_0)=-1$ MeV in
isospin-symmetric $\Lambda$-matter, a value close to that suggested by the
Nagara event \cite{Ahn2013}, using the following, 
\begin{equation} \label{eq:lambda-lambda-potential}
  U_{\Lambda}^{~\!(\Lambda)}(n_0) = 
	  g_{\omega \Lambda} \bar\omega_0
  + g_{\phi \Lambda} \bar\phi_0
  - g_{\sigma \Lambda} \bar\sigma_0
  - g_{\sigma^* \Lambda} \bar\sigma^*_0\,.
\end{equation}
The other strange-scalar meson-hyperon couplings are determined relative to
that of the $\Lambda$ as follows \cite{Oertel2015}, 
\begin{equation} \label{eq:xi-self-potential}
  U_{\Xi}^{(\Xi)}(n_0) = 2 U_{\Lambda}^{(\Lambda)}(n_0/2)\,,
\end{equation}
\begin{equation} \label{eq:strange-scalar-meson-Sigma-coupling}
  g_{\sigma^*\Sigma} = g_{\sigma^*\Lambda}\,.
\end{equation}

The isovector-vector meson-hyperon coupling constants $g_{\rho Y}$ are given as 
follows,
\begin{equation}
  g_{\rho\Lambda} = 0,\,\,\, g_{\rho\Sigma} = g_{\rho\Xi} = g_{\rho N}\,,
\end{equation}
with the differences in hyperon isospins accounted for by the isospin operator
in the Lagrangian (see Equation \eqref{eq:Lagrangian}). 


\section{Constraints from GW170817} \label{sec:gw170817}

\begin{figure}[t]
\centerline{\includegraphics[width=0.50\textwidth]{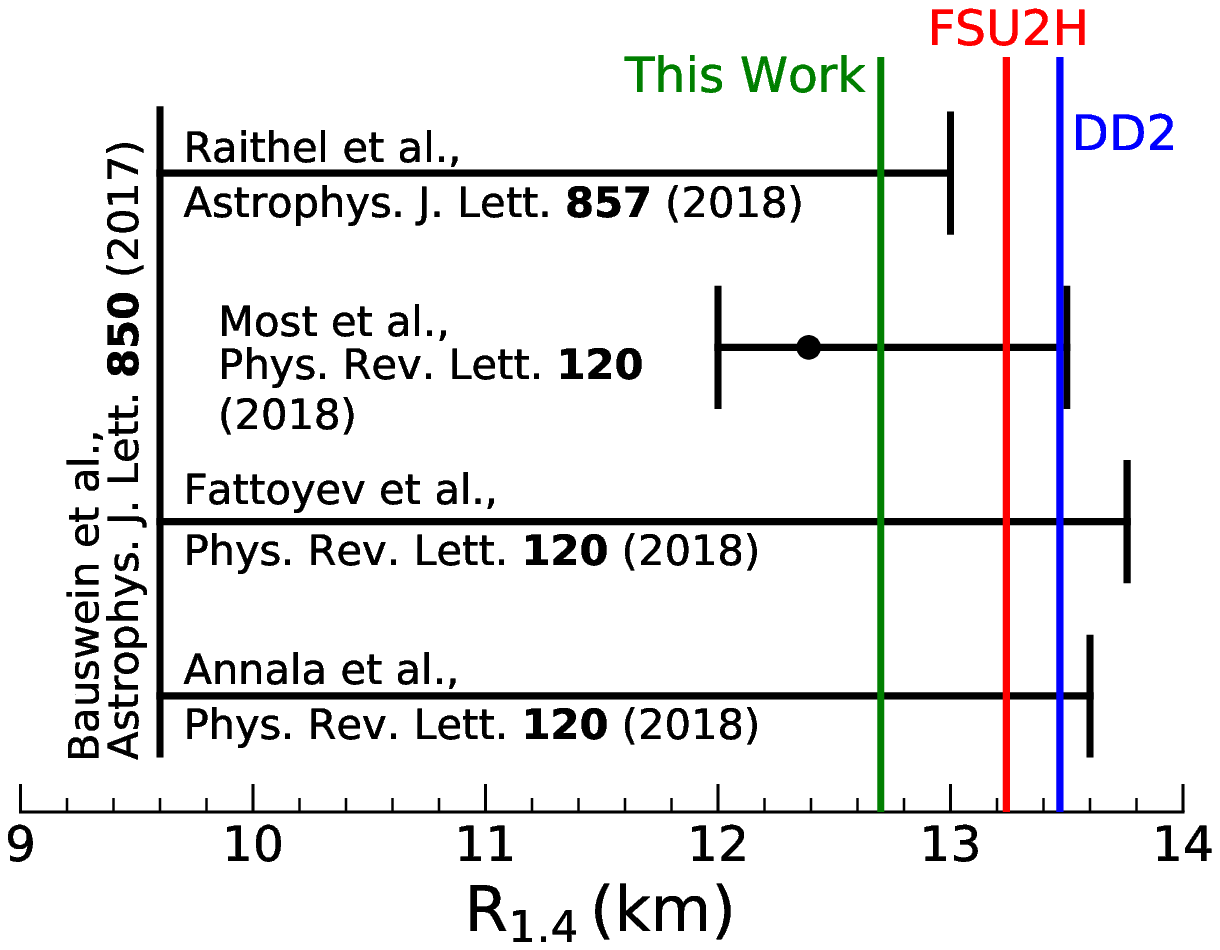}}
\caption{Constraints on the canonical NS radius derived using data
  from the recently observed BNS merger and gravitational wave signal GW170817
  \cite{Bauswein2017,Annala2018,Fattoyev2018,Most2018,Raithel2018}.}
\label{fig:canonical-radius}
\end{figure}
Analyses of gravitational waves (GW170817) and a gamma-ray burst
(GRB 170817A) emitted from the recent BNS merger have provided
additional constraints on the NS EoS, with particularly tight constraints
on hyperonic EoSs coming from estimates of the NS maximum mass
\cite{Abbott2017a,Abbott2017b}.
A range of $2.15\,M_{\odot} \le M_{\mathrm{max}} \le 2.25\,M_{\odot}$
was determined from numerical relativity simulations and
electromagnetic observations \cite{Shibata2017}.
Kilonova modeling suggests
$2.01\pm 0.04\,M_{\odot} \le M_{\mathrm{max}} \le 2.16^{+0.17}_{-0.15} \,M_{\odot}$,
the lower limit taken to be the measured mass of PSR J0348+0432
\cite{Rezzolla2018}.
Finally, analyses of the gamma-ray burst and kilonova ejecta provide an
upper limit of $M_{\mathrm{max}} \le 2.17\,M_{\odot}$ \cite{Margalit2017}.
These maximum mass constraints are also relatively consistent with the
measured mass of PSR J2215+5135.
From these findings, and for the purposes of constraining the hyperonic
EoSs calculated in this work, we determine a likely range for the maximum
NS mass to be
$2.01\,M_{\odot} \le M_{\mathrm{max}} \le 2.16\,M_{\odot}$
\cite{Banik2017}.

Several constraints on the canonical radius have also been deduced
from GW170817 and GRB 170817A and are presented in Figure
\ref{fig:canonical-radius}.
As shown, the EoS utilized in this work is consistent with these constraints
with a canonical radius of 12.7 km.
However, hyperons are not typically present in large enough quantities
in a $1.4\,M_{\odot}$ NS to have an appreciable affect on the radius, so
these constraints are not directly relevant to the hyperonic EoSs
presented in the remainder of this work.


\section{Results} \label{sec:results}

A thorough investigation of hyperonic NS EoSs requires the calculation of NS
properties for a large range of possible meson-hyperon coupling constants.
In this section we will discuss NS properties calculated for a range of 
vector meson-hyperon coupling constants specified by varying the $\alpha_V$
and $z$ SU(3) coupling parameters, with the scalar meson-hyperon couplings
fit to the hyperon saturation potentials as discussed in Section
\ref{sec:coupling-constants}.
To examine the affect of including hyperon-hyperon interactions,
we will compute NS properties for EoS models that include
the $\sigma$, $\omega$, and $\rho$ mesons ($\sigma\omega\rho$), the 
$\phi$ meson ($\sigma\omega\rho\phi$), and the $\sigma^*$ meson
($\sigma\omega\rho\phi\sigma^*$).
%
\begin{figure}[t]
\centerline{\includegraphics[width=0.50\textwidth]{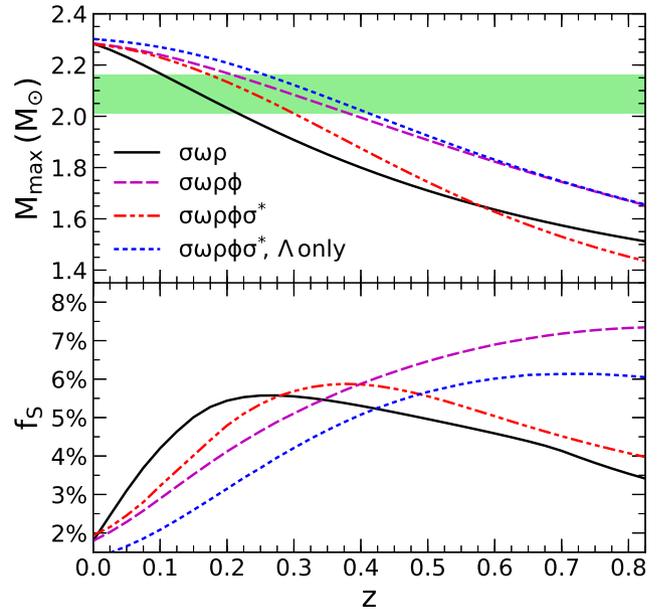}}
\caption{Maximum mass $M_{\mathrm{max}}$ (top panel) and strangeness fraction
$f_S$ (bottom panel) vs. $z$ for EoSs with the
inclusion of the indicated meson fields. The green highlighted region
(top panel) extends from $2.01-2.16\,M_{\odot}$, encapsulating
mass constraints from PSR J0348+0432 and GW170817.}
\label{fig:mass-strangeness-fraction}
\end{figure}
%
\begin{figure}[t]
\centerline{\includegraphics[width=0.50\textwidth]{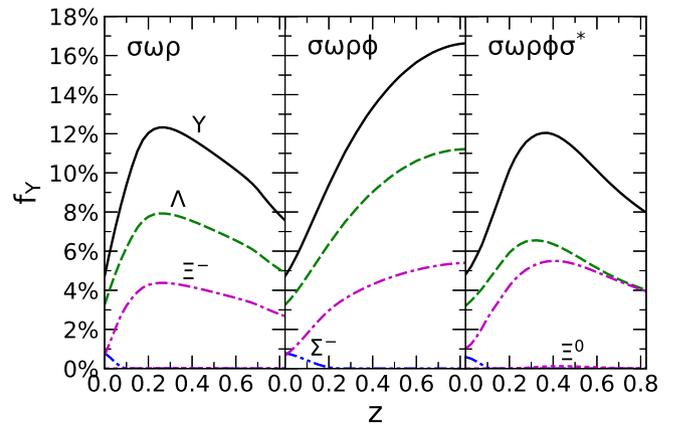}}
\caption{Maximum hyperon fractions $f_Y$ vs. $z$ for EoSs with the combinations
of mesons shown in the different panels. The solid black line labeled $Y$
represents the total hyperon fraction.}
\label{fig:hyperon-fraction}
\end{figure}

First we set $\alpha_V=1$ and calculate the maximum mass ($M_{\mathrm{max}}$)
and strangess fractions ($f_S$) (Figure \ref{fig:mass-strangeness-fraction})
as well as the maximum hyperon fractions ($f_Y$) (Figure \ref{fig:hyperon-fraction})
of NSs for all possible $z$.
Decreasing $z$ increases the values of the vector couplings $g_{\omega Y}$
and $|g_{\phi Y}|$, along with the maximum NS mass.
The $\sigma\omega\rho\phi$ model produces a stiff EoS with maximum masses up to
$\sim 0.2\,M_{\odot}$ greater than that of the $\sigma\omega\rho$ and
$\sigma\omega\rho\phi\sigma^*$ models.
However, if one considers the possibility that $\Sigma$ and more importantly
$\Xi$ hyperons do not appear in NS matter, then the maximum mass of the
$\sigma\omega\rho\phi\sigma^*$ model increases dramatically, the $\Xi-$ being
responsible for considerable softening of the EoS.
Total strangeness and hyperon fractions of the $\sigma\omega\rho\phi$ model
increase monotonically with $z$, while those of the 
$\sigma\omega\rho\phi\sigma^*$ model reach maximums
$f_{S}^{\mathrm{max}}\approx6\%$ and $f_{Y}^{\mathrm{max}}\approx12\%$
at $z\approx 0.37$ before decreasing with increasing $z$.

%
\begin{table}
  \begin{center}
    \begin{tabular}{cccc}
      \hline\noalign{\smallskip}
      Model & $z$ & $f_S$ & $f_Y$\\
      \noalign{\smallskip}\hline\noalign{\smallskip}
      $\sigma\omega\rho$ & 0.217 & 5.50\% & 12.2\% \\
      $\sigma\omega\rho\phi$ & 0.382 & 5.76\% & 13.0\% \\
      $\sigma\omega\rho\phi\sigma^*$ & 0.300 & 5.69\% & 11.8\% \\
      $\sigma\omega\rho\phi\sigma^*,\,\Lambda\,\mathrm{only}$ & 0.413 & 5.17\% & 15.5\% \\
      \noalign{\smallskip}\hline
    \end{tabular}
    \caption{Required $z$, maximum strangeness fractions $f_S$, and maximum
      hyperon fractions
      $f_Y$ for $2.01\,M_{\odot}$ maximum mass NSs extracted from the data in
      Figures~\ref{fig:mass-strangeness-fraction} and ~\ref{fig:hyperon-fraction}.}
    \label{table:model-properties-1}
  \end{center}
\end{table}
%
\begin{table}
  \begin{center}
    \begin{tabular}{cccc}
      \hline\noalign{\smallskip}
      Model & $z$ & $f_S$ & $f_Y$\\
      \noalign{\smallskip}\hline\noalign{\smallskip}
      $\sigma\omega\rho$ & 0.105 & 4.30\% & 9.58\% \\
      $\sigma\omega\rho\phi$ & 0.212 & 4.24\% & 9.67\% \\
      $\sigma\omega\rho\phi\sigma^*$ & 0.177 & 4.46\% & 9.58\% \\
      $\sigma\omega\rho\phi\sigma^*,\,\Lambda\,\mathrm{only}$ & 0.257 & 3.76\% & 11.2\% \\
      \noalign{\smallskip}\hline
    \end{tabular}
    \caption{Same as Table \ref{table:model-properties-1} but for
      $2.16\,M_{\odot}$ maximum mass NSs.}
    \label{table:model-properties-2}
  \end{center}
\end{table}

Results for the 2.01 $M_{\odot}$ and 2.16 $M_{\odot}$ mass constraint boundaries
extracted from Figures \ref{fig:mass-strangeness-fraction} and 
\ref{fig:hyperon-fraction} are summarized in Tables \ref{table:model-properties-1}
and \ref{table:model-properties-2}.
Only by excluding the $\Sigma$ and $\Xi$ hyperons from the
$\sigma\omega\rho\phi\sigma^*$ model can any of the calculated EoSs satisfy even
the $2.01\,M_{\odot}$ lower limit of the mass constraint, indicated by the bottom
of the green shaded region in Figure \ref{fig:mass-strangeness-fraction},
if SU(6) symmetry ($z_{\mathrm{SU(6)}}\approx0.408$) is assumed, let alone
the $2.16\,M_{\odot}$ upper limit.
If instead $z_{\mathrm{ESC08}} \approx 0.195$ is chosen, the 
$\sigma\omega\rho\phi$ and $\sigma\omega\rho\phi\sigma^*$ models easily satisfy
the $2.01\,M_{\odot}$ constraint with maximum masses
$M_{\mathrm{max}}(z_{\mathrm{ESC08}}) \approx 2.17\,M_{\odot}$ and
$M_{\mathrm{max}}(z_{\mathrm{ESC08}}) \approx 2.14\,M_{\odot}$ respectively.
However, the $\sigma\omega\rho\phi\sigma^*$ model requires $z\sim0.18$
(just under $z_{\mathrm{ESC08c}}\approx0.182$) to satisfy the $2.16\,M_{\odot}$
upper limit unless the $\Sigma$ and $\Xi$ hyperons are neglected, in which case
$z \sim 0.26$ is sufficient.

Hyperon fractions in $2.16\,M_{\odot}$ maximum mass NSs are substantial
at nearly 10\% when the entire baryon octet is included, with $f_{\Lambda}
= 5.6\%$ and $f_{\Xi^-}=3.9\%$ for the $\sigma\omega\rho\phi\sigma^*$
model.
The fraction of $\Lambda$s significantly exceeds that of the $\Xi^-$
in the $\sigma\omega\rho$ and $\sigma\omega\rho\phi$ models, but the two are
much more comparable when the $\sigma^*$ meson is included as shown.
Since no experimental data from the study of double-hypernuclei yet exists to
constrain the $\Xi$ saturation self-potential, $g_{\sigma^*\Xi}$ was instead
fixed theoretically using Equation \eqref{eq:xi-self-potential}.
An overestimation of $g_{\sigma^*\Xi}$ will likewise overestimate $f_{\Xi^-}$,
softening the hyperonic EoS.
Therefore, it is possible that the $M_{\mathrm{max}}$ vs. $z$ relation in 
Figure \ref{fig:mass-strangeness-fraction} for the
$\sigma\omega\rho\phi\sigma^*$ model may be closer to that displayed for the
$\Lambda$ hyperon alone, making the model more consistent with the stiff
mass constraints deduced from GW170817.
More experimental data from the study of hypernuclei and double-hypernuclei
is sorely needed to sufficiently constrain the hyperon single-particle-
and self-potentials.

\begin{figure}[t]
\centerline{\includegraphics[width=0.50\textwidth]{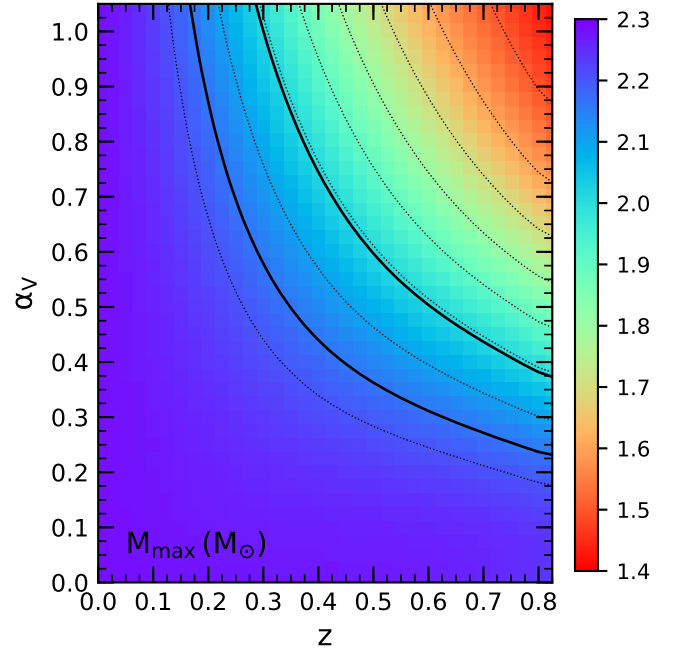}}
\caption{Maximum mass $M_{\mathrm{max}}\,(M_{\odot})$ in the $\alpha_V$ vs.
$z$ SU(3) parameter space for EoSs calculated using the
$\sigma\omega\rho\phi\sigma^*$ model.  Dotted lines are maximum mass contours
provided in $0.1\,M_{\odot}$ intervals, and solid lines represent the
$2.01\,M_{\odot}$ and $2.16\,M_{\odot}$ contours bounding the region
suggested by the mass of PSR J0348+0432 and analyses of GW170817.}
\label{fig:mass-heatmap}
\end{figure}
\begin{figure}[t]
\centerline{\includegraphics[width=0.50\textwidth]{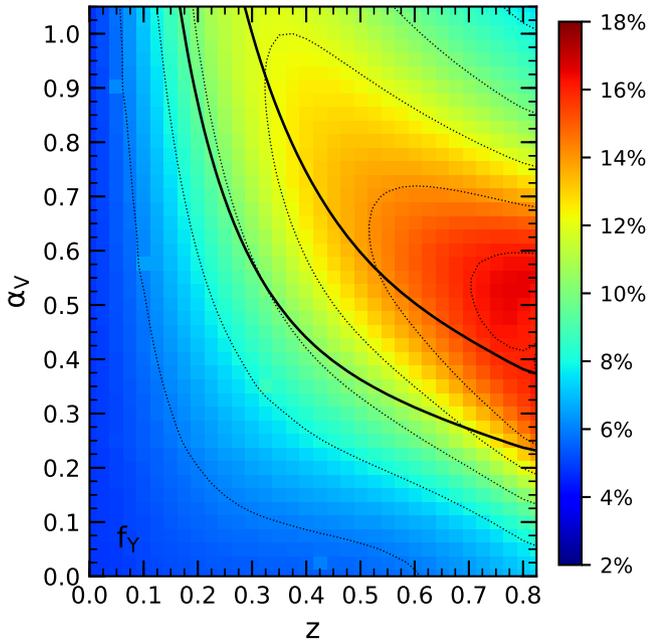}}
\caption{Hyperon fraction $f_Y$ in the $\alpha_V$ vs. $z$ SU(3) parameter space for
EoSs calculated using the $\sigma\omega\rho\phi\sigma^*$ model.
Dotted lines are hyperon fraction contours provided in $2\%$ intervals. The solid lines
represent the $2.01\,M_{\odot}$ and $2.16\,M_{\odot}$ contours bounding the region
suggested by the mass of PSR J0348+0432 and analyses of GW170817 as shown in Figure
\ref{fig:mass-heatmap}.}
\label{fig:hypfraction-heatmap}
\end{figure}

Finally, we let both the $\alpha_V$ and $z$ SU(3) parameters vary and calculate
the maximum NS mass and hyperon fraction for the entire space using the
$\sigma\omega\rho\phi\sigma^*$ model as shown in Figures \ref{fig:mass-heatmap}
and \ref{fig:hypfraction-heatmap}.
Neither $\alpha_V$ nor $z$ are necessarily restricted by the
$2.01-2.16\,M_{\odot}$ maximum mass constraint described by the bold
contours in Figure \ref{fig:mass-heatmap}.
Increasing $z$ beyond $\sim 0.3$ requires subsequent reduction of 
$\alpha_V$ below 1 to keep $M_{\mathrm{max}} > 2.01\,M_{\odot}$.
A hyperon fraction hotspot is centered around $\alpha_V = 0.525$ and
$z = 0.775$ with $f_Y^{\mathrm{max}} \approx 16.5\%$, while
hyperon fractions consistent with the mass constraints range from
$9.4-16\%$.


\section{Summary} \label{sec:summary}

In this contribution we introduced a new RMF parameterization, producing
NS EoSs consistent with tight constraints on the isospin asymmetry energy 
and slope, and with constraints on the canonical NS radius deduced from
analyses of the gravitational waves (GW170817) and gamma-ray burst
(GRB 170817A) emitted by the recent BNS merger.
This particular parameterization is of great utility in the study of
hyperonization, as it is relatively straightforward to recover the intended
saturation properties of symmetric nuclear matter when choosing vector
meson-hyperon coupling constants that are inconsistent with SU(6) symmetry
($z \ne 1/\sqrt{6}$).

We calculated the maximum mass, strangeness fraction, and hyperon fraction
of NSs for hyperonic EoSs, exploring the full range of possible vector
meson-hyperon coupling constants consistent with SU(3) symmetry.
Maximum masses were compared to constraints deduced from analyses
of GW170817 and GRB 170817A that suggest the maximum NS mass falls in
the following range:
$2.01\,M_{\odot} \le M_{\mathrm{max}} \le 2.16\,M_{\odot}$.
Taking the SU(3) parameter $\alpha_V=1$, we find this maximum mass
range suggests $0.18 \lesssim z \lesssim 0.30$ when considering the full
baryon octet and including both the $\phi$ and $\sigma^*$ mesons.
However, if only the $\Lambda$ hyperon is considered, we find that
$0.26 \lesssim z \lesssim 0.41$, highlighting the importance of
acquiring additional empirical data with which to constrain the saturation
single-particle- and self-potential of the $\Xi$ hyperon.
%
Finally, the strangeness and hyperon fractions found for the 
$\sigma\omega\rho\phi\sigma^*$ model and the given
maximum mass range are $4.5\% \lesssim f_S \lesssim 5.7\%$ and
$9.6\% \lesssim f_Y \lesssim 12\%$ respectively.

\section*{Acknowledgements}
We thank D. Blaschke for his valuable comments regarding this
manuscript.  This research was supported by the National
Science Foundation (USA), grant number PHY-1714068.

\bibliography{spinella_IWARA_2018}%

\begin{thebibliography}{}

\bibitem [\protect \citeauthoryear {%
Abbott%
\ \protect \BOthers {.}}{%
Abbott%
\ \protect \BOthers {.}}{%
{\protect \APACyear {2017}}%
{\protect \APACexlab {{\protect \BCnt {1}}}}}]{%
Abbott2017a}
\APACinsertmetastar {%
Abbott2017a}%
\begin{APACrefauthors}%
Abbott, B\BPBI P.%
\BCBT {}\ \BOthersPeriod {.}
\end{APACrefauthors}%
\unskip\
\newblock
\APACrefYearMonthDay{2017{\protect \BCnt {1}}}{}{},
\newblock
\unskip
\newblock
\APACjournalVolNumPages{Phys. Rev. Lett.}{119}{}{161101}.
\PrintBackRefs{\CurrentBib}

\bibitem [\protect \citeauthoryear {%
Abbott%
\ \protect \BOthers {.}}{%
Abbott%
\ \protect \BOthers {.}}{%
{\protect \APACyear {2017}}%
{\protect \APACexlab {{\protect \BCnt {2}}}}}]{%
Abbott2017b}
\APACinsertmetastar {%
Abbott2017b}%
\begin{APACrefauthors}%
Abbott, B\BPBI P.%
\BCBT {}\ \BOthersPeriod {.}
\end{APACrefauthors}%
\unskip\
\newblock
\APACrefYearMonthDay{2017{\protect \BCnt {2}}}{}{},
\newblock
\unskip
\newblock
\APACjournalVolNumPages{Astrophys. J. Lett.}{848}{}{L12}.
\PrintBackRefs{\CurrentBib}

\bibitem [\protect \citeauthoryear {%
Ahn%
\ \protect \BOthers {.}}{%
Ahn%
\ \protect \BOthers {.}}{%
{\protect \APACyear {2013}}%
}]{%
Ahn2013}
\APACinsertmetastar {%
Ahn2013}%
\begin{APACrefauthors}%
Ahn, J\BPBI K.%
\BCBT {}\ \BOthersPeriod {.}
\end{APACrefauthors}%
\unskip\
\newblock
\APACrefYearMonthDay{2013}{}{},
\newblock
\unskip
\newblock
\APACjournalVolNumPages{Phys. Rev. C}{88}{}{014003}.
\PrintBackRefs{\CurrentBib}

\bibitem [\protect \citeauthoryear {%
Annala%
, Gorda%
, Kurkela%
\BCBL {}\ \BBA {} Vuorinen%
}{%
Annala%
\ \protect \BOthers {.}}{%
{\protect \APACyear {2018}}%
}]{%
Annala2018}
\APACinsertmetastar {%
Annala2018}%
\begin{APACrefauthors}%
Annala, E.%
, Gorda, T.%
, Kurkela, A.%
\BCBL {}\ \BBA {} Vuorinen, A.%
\end{APACrefauthors}%
\unskip\
\newblock
\APACrefYearMonthDay{2018}{}{},
\newblock
\unskip
\newblock
\APACjournalVolNumPages{Phys. Rev. Lett.}{120}{}{172703}.
\PrintBackRefs{\CurrentBib}

\bibitem [\protect \citeauthoryear {%
Antoniadis%
\ \protect \BOthers {.}}{%
Antoniadis%
\ \protect \BOthers {.}}{%
{\protect \APACyear {2013}}%
}]{%
Antoniadis2013}
\APACinsertmetastar {%
Antoniadis2013}%
\begin{APACrefauthors}%
Antoniadis, J.%
\BCBT {}\ \BOthersPeriod {.}
\end{APACrefauthors}%
\unskip\
\newblock
\APACrefYearMonthDay{2013}{}{},
\newblock
\unskip
\newblock
\APACjournalVolNumPages{Science}{340}{}{6131}.
\PrintBackRefs{\CurrentBib}

\bibitem [\protect \citeauthoryear {%
{Banik}%
\ \BBA {} {Bandyopadhyay}%
}{%
{Banik}%
\ \BBA {} {Bandyopadhyay}%
}{%
{\protect \APACyear {2017}}%
}]{%
Banik2017}
\APACinsertmetastar {%
Banik2017}%
\begin{APACrefauthors}%
{Banik}, S.%
\BCBT {}\ \BBA {} {Bandyopadhyay}, D.%
\end{APACrefauthors}%
\unskip\
\newblock
\APACrefYearMonthDay{2017}{}{},
\newblock
\APACrefbtitle {{Dense Matter in Neutron Star: Lessons from GW170817}.} {{Dense
  Matter in Neutron Star: Lessons from GW170817}.}
\PrintBackRefs{\CurrentBib}

\bibitem [\protect \citeauthoryear {%
Bauswein%
, Just%
, Janka%
\BCBL {}\ \BBA {} Stergioulas%
}{%
Bauswein%
\ \protect \BOthers {.}}{%
{\protect \APACyear {2017}}%
}]{%
Bauswein2017}
\APACinsertmetastar {%
Bauswein2017}%
\begin{APACrefauthors}%
Bauswein, A.%
, Just, O.%
, Janka, H\BHBI T.%
\BCBL {}\ \BBA {} Stergioulas, A.%
\end{APACrefauthors}%
\unskip\
\newblock
\APACrefYearMonthDay{2017}{}{},
\newblock
\unskip
\newblock
\APACjournalVolNumPages{Astrophys. J. Lett.}{850}{}{L34}.
\PrintBackRefs{\CurrentBib}

\bibitem [\protect \citeauthoryear {%
Birkhan%
, Miorelli%
, Bacca%
\BCBL {}\ \protect \BOthers {.}}{%
Birkhan%
\ \protect \BOthers {.}}{%
{\protect \APACyear {2017}}%
}]{%
Birkhan2017}
\APACinsertmetastar {%
Birkhan2017}%
\begin{APACrefauthors}%
Birkhan, J.%
, Miorelli, M.%
, Bacca, S.%
\BCBL {}\ \BOthersPeriod {.}\end{APACrefauthors}%
\unskip\
\newblock
\APACrefYearMonthDay{2017}{}{},
\newblock
\unskip
\newblock
\APACjournalVolNumPages{Phys. Rev. Lett.}{118}{}{252501}.
\PrintBackRefs{\CurrentBib}

\bibitem [\protect \citeauthoryear {%
Danielewicz%
, Lacey%
\BCBL {}\ \BBA {} Lynch%
}{%
Danielewicz%
\ \protect \BOthers {.}}{%
{\protect \APACyear {2002}}%
}]{%
Danielewicz2002}
\APACinsertmetastar {%
Danielewicz2002}%
\begin{APACrefauthors}%
Danielewicz, P.%
, Lacey, R.%
\BCBL {}\ \BBA {} Lynch, W\BPBI G.%
\end{APACrefauthors}%
\unskip\
\newblock
\APACrefYearMonthDay{2002}{}{},
\newblock
\unskip
\newblock
\APACjournalVolNumPages{Science}{298}{}{1592}.
\PrintBackRefs{\CurrentBib}

\bibitem [\protect \citeauthoryear {%
Danielewicz%
\ \BBA {} Lee%
}{%
Danielewicz%
\ \BBA {} Lee%
}{%
{\protect \APACyear {2014}}%
}]{%
Danielewicz2014}
\APACinsertmetastar {%
Danielewicz2014}%
\begin{APACrefauthors}%
Danielewicz, P.%
\BCBT {}\ \BBA {} Lee, J.%
\end{APACrefauthors}%
\unskip\
\newblock
\APACrefYearMonthDay{2014}{}{},
\newblock
\unskip
\newblock
\APACjournalVolNumPages{Nucl. Phys. A}{922}{}{1}.
\PrintBackRefs{\CurrentBib}

\bibitem [\protect \citeauthoryear {%
Douchin%
\ \BBA {} Haensel%
}{%
Douchin%
\ \BBA {} Haensel%
}{%
{\protect \APACyear {2001}}%
}]{%
Douchin2001}
\APACinsertmetastar {%
Douchin2001}%
\begin{APACrefauthors}%
Douchin, F.%
\BCBT {}\ \BBA {} Haensel, P.%
\end{APACrefauthors}%
\unskip\
\newblock
\APACrefYearMonthDay{2001}{}{},
\newblock
\unskip
\newblock
\APACjournalVolNumPages{Astron. Astrophys.}{380}{}{151}.
\PrintBackRefs{\CurrentBib}

\bibitem [\protect \citeauthoryear {%
Dover%
\ \BBA {} Gal%
}{%
Dover%
\ \BBA {} Gal%
}{%
{\protect \APACyear {1984}}%
}]{%
Dover1984}
\APACinsertmetastar {%
Dover1984}%
\begin{APACrefauthors}%
Dover, C\BPBI B.%
\BCBT {}\ \BBA {} Gal, A.%
\end{APACrefauthors}%
\unskip\
\newblock
\APACrefYearMonthDay{1984}{}{},
\newblock
\unskip
\newblock
\APACjournalVolNumPages{Prog. Part. Nucl. Phys.}{12}{}{171}.
\PrintBackRefs{\CurrentBib}

\bibitem [\protect \citeauthoryear {%
Drago%
, Lavagno%
, Pagliara%
\BCBL {}\ \BBA {} Pigato%
}{%
Drago%
\ \protect \BOthers {.}}{%
{\protect \APACyear {2014}}%
}]{%
Drago2014a}
\APACinsertmetastar {%
Drago2014a}%
\begin{APACrefauthors}%
Drago, A.%
, Lavagno, A.%
, Pagliara, G.%
\BCBL {}\ \BBA {} Pigato, D.%
\end{APACrefauthors}%
\unskip\
\newblock
\APACrefYearMonthDay{2014}{}{},
\newblock
\unskip
\newblock
\APACjournalVolNumPages{Phys. Rev. C}{90}{}{065809}.
\PrintBackRefs{\CurrentBib}

\bibitem [\protect \citeauthoryear {%
Fattoyev%
, Piekarewicz%
\BCBL {}\ \BBA {} Horowitz%
}{%
Fattoyev%
\ \protect \BOthers {.}}{%
{\protect \APACyear {2018}}%
}]{%
Fattoyev2018}
\APACinsertmetastar {%
Fattoyev2018}%
\begin{APACrefauthors}%
Fattoyev, F\BHBI J.%
, Piekarewicz, J.%
\BCBL {}\ \BBA {} Horowitz, C\BPBI J.%
\end{APACrefauthors}%
\unskip\
\newblock
\APACrefYearMonthDay{2018}{}{},
\newblock
\unskip
\newblock
\APACjournalVolNumPages{Phys. Rev. Lett.}{120}{}{172702}.
\PrintBackRefs{\CurrentBib}

\bibitem [\protect \citeauthoryear {%
Fuchs%
}{%
Fuchs%
}{%
{\protect \APACyear {2006}}%
}]{%
Fuchs2006}
\APACinsertmetastar {%
Fuchs2006}%
\begin{APACrefauthors}%
Fuchs, C.%
\end{APACrefauthors}%
\unskip\
\newblock
\APACrefYearMonthDay{2006}{}{},
\newblock
\unskip
\newblock
\APACjournalVolNumPages{Prog. Part. Nucl. Phys.}{56}{}{1}.
\PrintBackRefs{\CurrentBib}

\bibitem [\protect \citeauthoryear {%
Haensel%
\ \BBA {} Pichon%
}{%
Haensel%
\ \BBA {} Pichon%
}{%
{\protect \APACyear {1994}}%
}]{%
Haensel1994}
\APACinsertmetastar {%
Haensel1994}%
\begin{APACrefauthors}%
Haensel, P.%
\BCBT {}\ \BBA {} Pichon, B.%
\end{APACrefauthors}%
\unskip\
\newblock
\APACrefYearMonthDay{1994}{}{},
\newblock
\unskip
\newblock
\APACjournalVolNumPages{Astron. Astrophys.}{283}{}{313}.
\PrintBackRefs{\CurrentBib}

\bibitem [\protect \citeauthoryear {%
Hagen%
\ \protect \BOthers {.}}{%
Hagen%
\ \protect \BOthers {.}}{%
{\protect \APACyear {2015}}%
}]{%
Hagen2015}
\APACinsertmetastar {%
Hagen2015}%
\begin{APACrefauthors}%
Hagen, G.%
\BCBT {}\ \BOthersPeriod {.}
\end{APACrefauthors}%
\unskip\
\newblock
\APACrefYearMonthDay{2015}{}{},
\newblock
\unskip
\newblock
\APACjournalVolNumPages{Nat. Phys.}{12}{}{186}.
\PrintBackRefs{\CurrentBib}

\bibitem [\protect \citeauthoryear {%
Hebeler%
, Lattimer%
, Pethick%
\BCBL {}\ \BBA {} Schwenk%
}{%
Hebeler%
\ \protect \BOthers {.}}{%
{\protect \APACyear {2013}}%
}]{%
Hebeler2013}
\APACinsertmetastar {%
Hebeler2013}%
\begin{APACrefauthors}%
Hebeler, K.%
, Lattimer, J\BPBI M.%
, Pethick, C\BPBI J.%
\BCBL {}\ \BBA {} Schwenk, A.%
\end{APACrefauthors}%
\unskip\
\newblock
\APACrefYearMonthDay{2013}{}{},
\newblock
\unskip
\newblock
\APACjournalVolNumPages{Astrophys. J.}{773}{}{11}.
\PrintBackRefs{\CurrentBib}

\bibitem [\protect \citeauthoryear {%
Lattimer%
\ \BBA {} Steiner%
}{%
Lattimer%
\ \BBA {} Steiner%
}{%
{\protect \APACyear {2014}}%
}]{%
Lattimer2014}
\APACinsertmetastar {%
Lattimer2014}%
\begin{APACrefauthors}%
Lattimer, J\BPBI M.%
\BCBT {}\ \BBA {} Steiner, A\BPBI W.%
\end{APACrefauthors}%
\unskip\
\newblock
\APACrefYearMonthDay{2014}{}{},
\newblock
\unskip
\newblock
\APACjournalVolNumPages{Eur. Phys. J. A}{50}{}{40}.
\PrintBackRefs{\CurrentBib}

\bibitem [\protect \citeauthoryear {%
Linares%
, Shahbaz%
\BCBL {}\ \BBA {} Casares%
}{%
Linares%
\ \protect \BOthers {.}}{%
{\protect \APACyear {2018}}%
}]{%
Linares2018}
\APACinsertmetastar {%
Linares2018}%
\begin{APACrefauthors}%
Linares, M.%
, Shahbaz, T.%
\BCBL {}\ \BBA {} Casares, J.%
\end{APACrefauthors}%
\unskip\
\newblock
\APACrefYearMonthDay{2018}{}{},
\newblock
\unskip
\newblock
\APACjournalVolNumPages{Astrophys. J.}{859}{}{54}.
\PrintBackRefs{\CurrentBib}

\bibitem [\protect \citeauthoryear {%
Lynch%
\ \protect \BOthers {.}}{%
Lynch%
\ \protect \BOthers {.}}{%
{\protect \APACyear {2009}}%
}]{%
Lynch2009}
\APACinsertmetastar {%
Lynch2009}%
\begin{APACrefauthors}%
Lynch, W\BPBI G.%
, Tsang, M\BPBI B.%
, Zhang, Y.%
, Danielewicz, P.%
, Famiano, M.%
, Li, Z.%
\BCBL {}\ \BBA {} Steiner, A\BPBI W.%
\end{APACrefauthors}%
\unskip\
\newblock
\APACrefYearMonthDay{2009}{}{},
\newblock
\unskip
\newblock
\APACjournalVolNumPages{Prog. Part. Nucl. Phys.}{62}{}{427}.
\PrintBackRefs{\CurrentBib}

\bibitem [\protect \citeauthoryear {%
Margalit%
\ \BBA {} Metzger%
}{%
Margalit%
\ \BBA {} Metzger%
}{%
{\protect \APACyear {2017}}%
}]{%
Margalit2017}
\APACinsertmetastar {%
Margalit2017}%
\begin{APACrefauthors}%
Margalit, B.%
\BCBT {}\ \BBA {} Metzger, B\BPBI D.%
\end{APACrefauthors}%
\unskip\
\newblock
\APACrefYearMonthDay{2017}{}{},
\newblock
\unskip
\newblock
\APACjournalVolNumPages{Astrophys. J. Lett.}{850}{}{L19}.
\PrintBackRefs{\CurrentBib}

\bibitem [\protect \citeauthoryear {%
Most%
, Weih%
, Rezzolla%
\BCBL {}\ \BBA {} Schaffner-Bielich%
}{%
Most%
\ \protect \BOthers {.}}{%
{\protect \APACyear {2018}}%
}]{%
Most2018}
\APACinsertmetastar {%
Most2018}%
\begin{APACrefauthors}%
Most, E\BPBI R.%
, Weih, L\BPBI R.%
, Rezzolla, L.%
\BCBL {}\ \BBA {} Schaffner-Bielich, J.%
\end{APACrefauthors}%
\unskip\
\newblock
\APACrefYearMonthDay{2018}{}{},
\newblock
\unskip
\newblock
\APACjournalVolNumPages{Phys. Rev. Lett.}{120}{}{261103}.
\PrintBackRefs{\CurrentBib}

\bibitem [\protect \citeauthoryear {%
Oertel%
, Providencia%
, Gulminelli%
\BCBL {}\ \BBA {} Raduta%
}{%
Oertel%
\ \protect \BOthers {.}}{%
{\protect \APACyear {2015}}%
}]{%
Oertel2015}
\APACinsertmetastar {%
Oertel2015}%
\begin{APACrefauthors}%
Oertel, M.%
, Providencia, C.%
, Gulminelli, F.%
\BCBL {}\ \BBA {} Raduta, A\BPBI R.%
\end{APACrefauthors}%
\unskip\
\newblock
\APACrefYearMonthDay{2015}{}{},
\newblock
\unskip
\newblock
\APACjournalVolNumPages{J. Phys. G}{42}{}{075202}.
\PrintBackRefs{\CurrentBib}

\bibitem [\protect \citeauthoryear {%
Raithel%
, Ozel%
\BCBL {}\ \BBA {} Psaltis%
}{%
Raithel%
\ \protect \BOthers {.}}{%
{\protect \APACyear {2018}}%
}]{%
Raithel2018}
\APACinsertmetastar {%
Raithel2018}%
\begin{APACrefauthors}%
Raithel, C\BPBI A.%
, Ozel, F.%
\BCBL {}\ \BBA {} Psaltis, D.%
\end{APACrefauthors}%
\unskip\
\newblock
\APACrefYearMonthDay{2018}{}{},
\newblock
\unskip
\newblock
\APACjournalVolNumPages{Astrophys. J. Lett.}{857}{}{L23}.
\PrintBackRefs{\CurrentBib}

\bibitem [\protect \citeauthoryear {%
Rezzolla%
, Most%
\BCBL {}\ \BBA {} Weih%
}{%
Rezzolla%
\ \protect \BOthers {.}}{%
{\protect \APACyear {2018}}%
}]{%
Rezzolla2018}
\APACinsertmetastar {%
Rezzolla2018}%
\begin{APACrefauthors}%
Rezzolla, L.%
, Most, E\BPBI R.%
\BCBL {}\ \BBA {} Weih, L\BPBI R.%
\end{APACrefauthors}%
\unskip\
\newblock
\APACrefYearMonthDay{2018}{}{},
\newblock
\unskip
\newblock
\APACjournalVolNumPages{Astrophys. J. Lett.}{852}{}{L25}.
\PrintBackRefs{\CurrentBib}

\bibitem [\protect \citeauthoryear {%
Rijken%
, Nagels%
\BCBL {}\ \BBA {} Yamamoto%
}{%
Rijken%
\ \protect \BOthers {.}}{%
{\protect \APACyear {2010}}%
}]{%
Rijken2010}
\APACinsertmetastar {%
Rijken2010}%
\begin{APACrefauthors}%
Rijken, T\BPBI A.%
, Nagels, M\BPBI M.%
\BCBL {}\ \BBA {} Yamamoto, Y.%
\end{APACrefauthors}%
\unskip\
\newblock
\APACrefYearMonthDay{2010}{}{},
\newblock
\unskip
\newblock
\APACjournalVolNumPages{Prog. Theor. Phys. Suppl.}{185}{}{14}.
\PrintBackRefs{\CurrentBib}

\bibitem [\protect \citeauthoryear {%
Roca-Maza%
\ \protect \BOthers {.}}{%
Roca-Maza%
\ \protect \BOthers {.}}{%
{\protect \APACyear {2015}}%
}]{%
Roca-Maza2015}
\APACinsertmetastar {%
Roca-Maza2015}%
\begin{APACrefauthors}%
Roca-Maza, X.%
, Vinas, X.%
, Centelles, M.%
\ et al.\end{APACrefauthors}%
\unskip\
\newblock
\APACrefYearMonthDay{2015}{}{},
\newblock
\unskip
\newblock
\APACjournalVolNumPages{Phys. Rev. C}{92}{}{064304}.
\PrintBackRefs{\CurrentBib}

\bibitem [\protect \citeauthoryear {%
Shibata%
\ \protect \BOthers {.}}{%
Shibata%
\ \protect \BOthers {.}}{%
{\protect \APACyear {2017}}%
}]{%
Shibata2017}
\APACinsertmetastar {%
Shibata2017}%
\begin{APACrefauthors}%
Shibata, M.%
, Fujibayashi, S.%
, Hotokezaka, K.%
, Kiuchi, K.%
, Kyutoku, K.%
, Sekiguchi, Y.%
\BCBL {}\ \BBA {} M., T.%
\end{APACrefauthors}%
\unskip\
\newblock
\APACrefYearMonthDay{2017}{}{},
\newblock
\unskip
\newblock
\APACjournalVolNumPages{Phys. Rev. C}{96}{}{123012}.
\PrintBackRefs{\CurrentBib}

\bibitem [\protect \citeauthoryear {%
Tews%
, Lattimer%
, Ohnishi%
\BCBL {}\ \BBA {} Kolomeitsev%
}{%
Tews%
\ \protect \BOthers {.}}{%
{\protect \APACyear {2017}}%
}]{%
Tews2017}
\APACinsertmetastar {%
Tews2017}%
\begin{APACrefauthors}%
Tews, I.%
, Lattimer, J\BPBI M.%
, Ohnishi, A.%
\BCBL {}\ \BBA {} Kolomeitsev, E\BPBI E.%
\end{APACrefauthors}%
\unskip\
\newblock
\APACrefYearMonthDay{2017}{}{},
\newblock
\unskip
\newblock
\APACjournalVolNumPages{Astrophys. J.}{848}{}{105}.
\PrintBackRefs{\CurrentBib}

\bibitem [\protect \citeauthoryear {%
Typel%
\ \BBA {} Wolter%
}{%
Typel%
\ \BBA {} Wolter%
}{%
{\protect \APACyear {1999}}%
}]{%
Typel1999}
\APACinsertmetastar {%
Typel1999}%
\begin{APACrefauthors}%
Typel, S.%
\BCBT {}\ \BBA {} Wolter, H\BPBI H.%
\end{APACrefauthors}%
\unskip\
\newblock
\APACrefYearMonthDay{1999}{}{},
\newblock
\unskip
\newblock
\APACjournalVolNumPages{Nucl. Phys. A}{656}{}{331}.
\PrintBackRefs{\CurrentBib}

\end{thebibliography}

\end{document}